**Behavioral Finance -- Asset Prices Predictability, Equity Premium Puzzle, Volatility Puzzle: The Rational Finance Approach**

**Svetlozar Rachev**
Texas Tech University

**Stoyan Stoyanov**
Stony Brook University

**Stefan Mittnik**
Technical University, Berlin

**Frank J. Fabozzi**
EDHEC Business School

**Abootaleb Shirvani**
Texas Tech University

**Abstract:** In this paper we address three main objections of behavioral finance to the theory of rational finance, considered as "anomalies" the theory of rational finance cannot explain: (i) Predictability of asset returns; (ii) The Equity Premium; (iii) The Volatility Puzzle. We offer resolutions of those objections within the rational finance. We do not claim that those are the only possible explanations of the "anomalies", but offer statistical models within the rational theory of finance which can be used without relying on behavioral finance assumptions when searching for explanations of those "anomalies".

## 1.**Introduction**

In1995, economist Werner De Bondt wrote*," The sad truth is that modern finance theory offers only a set of asset pricing models for which little support exists and a set of empirical facts for which no theory exists".* (p. 8). Even economist, Nobel Laureate, and standard finance pioneer Merton Miller admitted, in an April, 23, 1994, interview with the Economist that conventional economics had failed to explain how asset prices are set. He added, however, that he believed the new mix of psychology and finance would lead nowhere.

Richard Zechhauser (1998) (p. 436) wrote: “*I do not think that the conflict between rationalists and behavioralists will be resolved in an intellectual generation, or even 3 such generations. There are simply too many battlefields. Each side can select the ones most favorable to its own cause. From time to time there will be mutually agreed-on skirmishes. Major recent ones have centered on macroeconomics, where the evidence remains exceedingly controversial and inclusive, and finance, where markets work exceedingly well but not perfectly-an outcome, I suspect, the behavioralists will continue mounting experiments or micro evidence of non-rational chooses, for there are infinite number to be found. The rationalists will take succor from the overwhelmingly power of their model, which had a lot to do with its success in the first place, and the absence of any equivalently power competitor. Should behavior in certain salient areas be found to violate the rationality, it will be treated as beyond economics. Decisions on religion and, conceivably, on family choices or personal habits thus may command the rationalists’ attention, if they behave well: otherwise, they may be classified in the same category as the source of preferences or values, something about which we have little to add as economics.”*

Statman (2014) asserted: *“Behavioral finance is under construction as a solid structure of finance. It incorporates parts of standard finance, replaces others, and includes bridges between theory, evidence, and practice.” Behavioral finance substitutes normal people for the rational people in standard finance. It substitutes behavioral portfolio theory for mean-variance portfolio theory, and behavioral asset pricing model for the CAPM and other models where expected returns are determined only by risk. Behavioral finance also distinguishes rational markets from hard-to-beat markets in the discussion of efficient markets, a distinction that is often blurred in standard finance, and it examines why so many investors believe that it is easy to beat the market. Moreover, behavioral finance expands the domain of finance beyond portfolios, asset pricing, and market* *efficiency and is set to continue that expansion while adhering to the scientific rigor introduced by standard finance."*

Our strong option is that **there is no scientific claim in the theory of behavioral finance, that could not be explained in a rational finance framework**. In this paper we a address the three main objections of behavioral finance proponents against the rational finance. $(i)$ Predictability of asset returns; $(ii)$The Equity Premium; $(iii)$ The Volatility Puzzle.[1]

---

[1] See for example Barberis and Thales (2003) who wrote: "Behavioral finance is a new approach to financial markets that has emerged, at least in part, in response to the difficulties faced by the traditional paradigm. In broad terms, it argues that some financial phenomena can be better understood using models in which some agents are not fully rational. More specifically, it analyzes what happens when we relax one, or both, of the two tenets that underlie individual rationality…. **4. Application: The aggregate stock market** Researchers studying the aggregate U.S. stock market have identified a number of interesting facts about its behavior. Three of the most striking are: **The Equity Premium**. The stock market has historically earned a high excess rate of return. For example, using annual data from 1871–1993, Campbell and Cochrane (1999) report that the average log return on the S&P 500 index is 3.9% higher than the average log return on short-term commercial paper. **Volatility**. Stock returns and price–dividend ratios are both highly variable. In the same data set, the annual standard deviation of excess log returns on the S&P 500 is 18%, while the annual standard deviation of the log price–dividend ratio is 0.27. An early discussion of this aversion can be found in Knight (1921), who defines risk as a gamble with known distribution and uncertainty as a gamble with unknown distribution, and suggests that people dislike uncertainty more than risk.

## 2. **The Predictability of Asset pricing: The Rational Finance Approach**

A major issue raised by the proponents of behavioral finance is that prices are often predictable[2], more precisely, given a stochastic basis $(\Omega, \mathcal{F}, \mathbb{F} = (\mathcal{F}_t, t \geq 0), \mathbb{P})$ a price process $S(t), t \geq 0$, defined on $(\Omega, \mathcal{F}, \mathbb{P})$ is not necessarily $\mathbb{F}$-adapted, it is adapted to an augmented filtration $\mathbb{F}^{(*)} \supset \mathbb{F}$, with $\mathbb{F}^{(*)} \subset \mathcal{F}$. The majority of work on predictability of asset returns is based on statistical, macro and fundamental and factor analyses[3]: (i) Conditional Capital Asset Prcing Model (CAPM); (ii) vector autoregressive (VAR) models; (iii) Bayesian statistical factor analysis; (iv) posterior moments of the predictable regression coefficients; (v) posterior odds; (vi) the information in asset prices; (vii), business cycles effects; (viii) asset predictability of future returns from initial dividend yields; (ix) firm characteristics as stock return predictors; (x) anomalies; (xi) predictive power of scaled-price ratios such as book-to market and earnings-to-price, forward spread, and short rate; (xii) variance risk premia and variance spillovers; (xiii) momentum ,market memory and reversals; (xiv) early announcements and others. Behavioral factors leading to financial assets

---

**Predictability**. Stock returns are forecastable. Using monthly, real, equal-weighted NYSE returns from 1941–1986, Fama and French (1988) show that the dividend– price ratio is able to explain 27% of the variation of cumulative stock returns over the subsequent four years.

[2] See for example, Daniel and Hirshleifer (2015): “Moreover, asset prices display patterns of predictability that are difficult to reconcile with rational expectations–based theories of price formation.

[3] See for example, Kandel and Stambaugh (1996), Neely and Weller (2000), Malkiel B.G. (2003), Barberis and Thaler (2003) , Shiller (2003), Avramov (2004), Wachter and Warusawitharana (2009) , Pesaran (2010), Zhou (2010), Bekiros (2013)

predictability[4]: (i) sentiment; (ii) overconfidence; (iii) optimism and wishful thinking; (iv) conservatism; euphoria and gloom; (v) self-deception; (vi) cursedness; (vii) belief perseverance; (viii) anchoring, etc.

Admitting the fact that asset returns are predictable Lo and Wang (1995) implement option pricing model when the asset returns are predictable. The model is based on specially designed Multivariate Trending O-U process, which are cumbersome, include many parameters, and those with small dimensions such univariate and bivariate trending O-U processes are not realistic as claimed in Lo and Wang (1995), pages 93 and 108. Our option pricing method is close to the Shiller's idea [5]of "smart money versus ordinary investors".

To model the predictability of asset prices, we use Stratonovich integral[6]:

$$\int_0^T \theta(t) \circ^{\left(\frac{1}{2}\right)} dB(t) =$$

$$= \lim_{0=t^{(0)}<t^{(1)}<\cdots<t^{(k)}=T,\, t^{(j)}=j\Delta t,\, \Delta t\downarrow 0} \sum_{j=0}^{k-1} \theta\left(\frac{t^{(j+1)}+t^{(j)}}{2}\right)\left(B(t^{(j+1)}) - B(t^{(j)})\right). \qquad (1)$$

In (1), $B(t), t \geq 0,$ is a Brownian motion generating a stochastic basis $(\Omega, \mathcal{F}, \mathbb{F} = (\mathcal{F}_t, t \geq 0), \mathbb{P})$, $\theta(t), t \geq 0$ is $\mathbb{F}$-adapted left-continuous and locally bonded process. The convergence is understood in probability. Important property is that Stratonovich integral "looks into the future",

---

[4] See Lewellen (2000), Hirshleifer (2001) , Barberis and Thaler (2003), Ferson (2006),Peleg (2008), Chapter 1, Daniel and Hirshleifer (2015).

[5] Shiller (2003).

[6] See Kloeden, Platen and Schurz H. (2000) Chapter 2, Øksendal (2003) Chapter 5, Syga (2015)

and thus price processes based on Stratonovich integral possess predictability properties. In sharp contrast, the Itô integral:

$$\int_0^T \theta(t)dB(t) = \lim_{0=t^{(0)}<t^{(1)}<\cdots<t^{(k)}=T, t^{(j)}=j\Delta t, \Delta t\downarrow 0} \sum_{j=0}^{k-1} \theta\left(t^{(j)}\right)\left(B(t^{(j+1)}) - B(t^{(j)})\right) \qquad (2)$$

"does not look in the future", and thus Itô prices are not predictable. We combine both integrals (1) and (2) within an Stratonovich $\alpha$ -integral) with $\alpha \in [0,1]$:

$$\int_0^T \theta(t) \circ^{(\alpha)} dB(t) =$$

$$= \lim_{0=t^{(0)}<t^{(1)}<\cdots<t^{(k)}=T, t^{(j)}=j\Delta t, \Delta t\downarrow 0} \sum_{j=0}^{k-1} \theta\left(t^{(j)}(1-\alpha) + \alpha t^{(j+1)}\right)\left(B(t^{(j+1)}) - B(t^{(j)})\right) =$$

$$= 2\alpha \int_0^T \theta(t) \circ^{\left(\frac{1}{2}\right)} dB(t) + (1-2\alpha)\int_0^T \theta(t)dB(t). \qquad (3)$$

Consider market with two assets:

$(i)$ risky asset (stock) $\mathcal{S}$ with potentially predictive price process $S(t), t \geq 0,$ following Stratonovich $\alpha$-SDE:

$$dS(t) = \mu\left(t, S(t)\right)dt + \sigma\left(t, S(t)\right) \circ^{(\alpha)} dB(t), t \geq 0, S(0) > 0, \qquad (4)$$

for some $\alpha \in [0,1]$, that is,

$$dS(t) = \left(\mu\left(t, S(t)\right) + 2\alpha^2\sigma\left(t, S(t)\right) \frac{\partial\sigma\left(t, S(t)\right)}{\partial x}\right)dt + \sigma\left(t, S(t)\right)dB(t), t \geq 0. \qquad (5)$$

$(ii)$ riskless asset (bond) $\mathcal{B}$ with price process $\beta(t), t \geq 0,$ defined by

$$d\beta(t) = r\left(t, S(t)\right)\beta(t), \beta(0) > 0. \qquad (6)$$

Consider an European Contingent Claim (ECC) $\mathfrak{C}$ with price process $C(t) = C\left(t, S(t)\right)$,where $C(t,x), t \geq 0, x > 0$, has continuous derivatives $\frac{\partial C(t,x)}{\partial t}$ and $\frac{\partial^2 C(t,x)}{\partial x^2}$. Assume that a trader ℶ takes a short position in $\mathfrak{C}$. Furthermore, when ℶ trades stock $\mathcal{S}$, with possibly superior or inferior to (4) dynamics, following Stratonovich $\gamma$-SDE:

$$dS(t) = \mu\big(t, S(t)\big)dt + \sigma\big(t, S(t)\big) \circ^{(\gamma)} dB(t), t \geq 0, S(0) > 0, \tag{4}$$

for some $\gamma \in [0,1]$, that is,

$$dS(t) = \left(\mu\big(t, S(t)\big) + 2\gamma^2\sigma\big(t, S(t)\big) \frac{\partial\sigma\big(t,S(t)\big)}{\partial x}\right)dt + \sigma\big(t, S(t)\big)dB(t), t \geq 0. \tag{5}$$

By the Itô formula:

$$dC\big(t, S(t)\big) =$$

$$= \left\{\frac{\partial C\big(t,S(t)\big)}{\partial t} + \frac{\partial C\big(t,S(t)\big)}{\partial x}\left(\mu\big(t,S(t)\big) + 2\alpha^2\sigma\big(t,S(t)\big)\frac{\partial\sigma\big(t,S(t)\big)}{\partial x}\right)\right.$$

$$\left. + \frac{1}{2}\frac{\partial^2 C\big(t,S(t)\big)}{\partial x^2}\Big(\sigma\big(t,S(t)\big)\Big)^2\right\}dt + \frac{\partial C\big(t,S(t)\big)}{\partial x}\sigma\big(t,S(t)\big)dB(t).$$

ℶ$'s$ replicating self-financing strategy is given by the pair $a(t), b(t), t \geq 0$, where

$C\big(t, S(t)\big) = a(t)S(t) + b(t)\beta(t)$ with $dC\big(t, S(t)\big) = a(t)dS(t) + b(t)d\beta(t)$. Thus,

$$dC\big(t, S(t)\big) = \left(a(t)\left(\mu\big(t, S(t)\big) + 2\gamma^2\sigma\big(t, S(t)\big)\frac{\partial\sigma\big(t, S(t)\big)}{\partial x}\right) + b(t)r(t, S(t))\beta(t)\right)dt +$$

$$+a(t)\sigma\big(t, S(t)\big)dB(t).$$

Equating the terms with $dC\big(t, S(t)\big)$, leads to $a(t) = \frac{\partial C(t,S(t))}{\partial x}$, and $b(t)\beta(t) = C\big(t, S(t)\big) - \frac{\partial C(t,S(t))}{\partial x}S(t)$. Equating the terms with $dt$ and setting $S(t) = x$, results in the following PDE:

$$\frac{\partial C(t,x)}{\partial t} + \frac{\partial C(t,x)}{\partial x}\left(r(t,x)x + 2(\alpha^2-\gamma^2)\sigma(t,x)\;\frac{\partial \sigma(t,x)}{\partial x}\right) -$$

$$-r(t,x)C(t,x) + \frac{1}{2}\frac{\partial^2 C(t,x)}{\partial x^2}\left(\sigma(t,x)\right)^2 = 0. \qquad (6)$$

Denote by $v\big(t,S(t)\big), t \geq 0$, stock volatility $v\big(t,S(t)\big) = \frac{\sigma(t,S(t))}{S(t)} > 0, t \geq 0$. We assume that $v\big(t,S(t)\big) > 0, \frac{\partial v(t,S(t))}{\partial x} \geq 0$. We call $\wp \coloneqq \gamma - \alpha \in R$ *excess predictability of the stock traded by* ב. Denote by $D_y(t,x) = 2\wp(\gamma+\alpha)v(t,x)\left(\frac{\partial v(t,xx)}{\partial x}x + v(t,x)\right)$ "*dividend yield due to predictability*". Then, the PDE (6) becomes

$$\frac{\partial C(t,x)}{\partial t} + \frac{\partial C(t,x)}{\partial x}\Big(r(t,x) - D_y(t,x)\Big)x - r(t,x)C(t,x) + \frac{1}{2}\frac{\partial^2 C(t,x)}{\partial x^2}\big(v(t,x)\big)^2 x^2 = 0.$$

Depending on the sign of $\wp$ $D_y(t,x)$ could be positive or negative. Indeed when $\wp = 0$, we obtain the classical Black-Scholes equation.

Thus, we were able to incorporate the option pricing markets in markets with predictable asset returns within the rational Black-Scholes framework. In particular, $\mathcal{S}$-price dynamics is given by[7]

$$S(t) = \mathbb{E}_t^{\mathbb{Q}}\left\{e^{-\int_t^T r(u,S(u)du}S(T) + \int_t^T e^{-\int_t^s r(u,S(u)du}\,D_y(s,S(s))ds\right\},$$

[7] See Duffie (2001), Section 6L.

where $\mathbb{Q}$ is equivalent martingale measure for dividend-stock-price pair $(D(t), X(t))$[8], where

$dX(t) = \mu\big(t, X(t)\big)dt + \sigma\big(t, X(t)\big)dB(t)$ and $D(t) = \int_0^t D_y(s, X(s))\, ds$.

As an application of our method consider the Heath-Jarrow-Morton (HJM)[9] model for the Term Structure of Interest Rates (TSIR) assuming predictability of the forward rates $f(t,u), 0 \leq t \leq u$. We assume that for every fixed $T \in (0, \mathcal{T}]$, $g(t,T) = e^{f(t,T)}, 0 \leq t < T$, satisfies

$$\frac{dg(t,T)}{g(t,T)} = \left(m(t,T) + \frac{1}{2}\big(v(t,T)\big)^2\right)dt + v(t,T) \circ^{(\alpha)} dB(t)$$

for some $\alpha \in [0,1]$, that is,

$$df(t,T) = (m(t,T) + 2\alpha^2 v(t,T)^2)dt + v(t,T)dB(t), t \geq 0.$$

Then, the no-arbitrage condition is that the market price of risk $\theta(t), t \geq 0$, given by

$$\theta(t) = \frac{m(t,T) + \; 2\alpha^2 v(t,T)^2 - v(t,T)\int_t^T v(t,u)\, du}{v(t,T)}$$

does not depend on the maturity $T$. Thus, under the risk-neutral measure $\mathbb{Q}\sim\mathbb{P}$, with $dB^{\mathbb{Q}}(t) = dB(t) + \theta(t)dt$, the risk-neutral dynamics of the forward rates is given by

$$df(t,T) = v(t,T)\left(\int_t^T v(t,u)\, du\right)dt + v(t,T)dB^{\mathbb{Q}}(t).$$

[8] That is, $\mathbb{Q}\sim\mathbb{P}$, and the discounted gain process $G^{(Y)}(t) = \; X^{(Y)}(t) + D^{(Y)}(t)$ is a $\mathbb{Q}$-martingale, where $Y(t) = \frac{1}{\beta(t)}, t \geq 0$, and $X^{(Y)}(t) = X(t)Y(t)$ and $dD^{(Y)}(t) = Y(t)dD(t)$.

[9] Heath, Jarrow, and Morton (1992), Brigo and Mercurio (2007), Chapter 5.

Then the risk-neutral dynamics $B(t,T), 0 \leq t < T,$ of the zero-coupon bond with maturity $T$ is determined by

$$\frac{dB(t,T)}{B(t,T)} = f(t,t) - v(t,T)\left(\int_t^T v(t,u)\,du\right)dB^{\mathbb{Q}}(t).$$

This example shows a possible approach to bond pricing when the interest rates or their derivatives are predictable.

## 3. **Equity Premium Puzzle: The Rational Finance Approach**

Mehra and Prescott (1985) introduced the equity premium puzzle[10], claiming that the historic equity premium in the US for the period 1889-1978, is not in line with the traditional asset pricing model, based on expected utility theory. The authors concluded that investors were much more risk averse than the traditional models would assume. Benartzi and Thaler (1995) suggested that "narrow framing" [11] leads investors to overestimate equity risk, and proposed an alternative to the standard investor preferences approach in Mehra and Prescott (1985), the so-called myopic loss aversion model, based on prospect theory[12]. It is based on experimental studies of human decisions under risk, rather than relying on assumption of purely rational market participants[13].

[10] The most extensive overview of the approaches to the equity premium puzzle is provided in Mehra (2008), for more recent works, we refer to Edelstein and Magin K. (2013), Cover and Zhuang (2016), Chen (2016) Kashyap (2016), Kliber (2016), Soklakov (2016), Tamura and Matsubayashi, ,McPherron (2017).

[11] See Millar (2013), Nada (2013), Zervoudi and Spyrou (2016).

Mehra and Prescott (1985) assumed that the growth rate of consumption and the dividends are log-normally distributed. Rietz (1988) claimed to resolve the equity-premium puzzle by assuming the possibilities for low-probability disastrous. His work was the starting point of the debate whether rare events can explain the Equity Premium puzzle, and the debate is still on[14].

In this section, we extend Mehra and Prescott (1985) approach to accommodate for rare events, by assuming that the growth rate of consumption and the dividends could be heavy-tailed distributed. In fact, we allow for a large spectrum of distributional tails, so that the statistical analysis of the data depending on the time period or country can determine the type of the distribution. What we can claim is that the distribution[15] of the growth rate of consumption is highly unlikely to be log-normal[16]. Much more flexible class of distribution is needed when modelling the growth rate. To this end, we suggest the Normal-Inverse Gaussian (NIG) distribution introduced by Ole Barndorff-Nielsen [17].

---

[12] Læssøe and Diernisse (2011)

[13] Andries (2012) incorporated loss aversion features in a recursive model of preferences and found tractable solutions to the consumption based asset pricing model with homogenous agents.

[14] Barro (2005), Jobert, Platania and Rogers (2006), Julliard and Ghosh (2012),

[15] $X$ is log-normally distributed, $X \triangleq ln\mathcal{N}(\mu,\sigma^2)$, denoted if $lnX$ is normally distributed $lnX \triangleq \mathcal{N}(\mu,\sigma^2)$ with mean $\mu$ and variance $\sigma^2$.

[16] See for example, Cont (2000), Bamberg and Neuhierl (2011),. Schmidt et al. (2012)

Random variable $X$ has NIG distribution, denoted $X \triangleq NIG(\mu,\alpha,\beta,\delta)$, $\mu \in R, \alpha \in R, \beta \in R, \delta \in R, \alpha^2 > \beta^2$if its density is given by

$$f_X(x) = \frac{\alpha\delta K_1(\alpha\sqrt{\delta^2+(x-\mu)^2}}{\pi\sqrt{\delta^2+(x-\mu)^2}} \exp\left\{\delta\sqrt{\alpha^2-\beta^2}+\beta(x-\mu)\right\}, x \in R.$$

Then, $X$ has mean $\mathbb{E}X = \mu + \frac{\delta\beta}{\sqrt{\alpha^2-\beta^2}}$, variance $varX = \frac{\delta\alpha^2}{(\alpha^2-\beta^2)^{\frac{3}{2}}}$, skewness $\gamma(X) = \frac{3\beta}{\alpha\sqrt{\delta}(\alpha^2-\beta^2)^{\frac{1}{4}}}$ and excess kurtosis $\kappa(X) = \frac{3\left(1+\frac{4\beta^2}{\alpha^2}\right)}{\delta(\alpha^2-\beta^2)^{\frac{1}{2}}}$. The characteristic function $\varphi_X(t) = \mathbb{E}e^{itX}, t \in R$, is given by

$$\varphi_X(t) = \exp\left\{i\mu t + \delta\left(\sqrt{\alpha^2-\beta^2} - \sqrt{\alpha^2-(\beta+it)^2}\right)\right\} \quad (7)$$

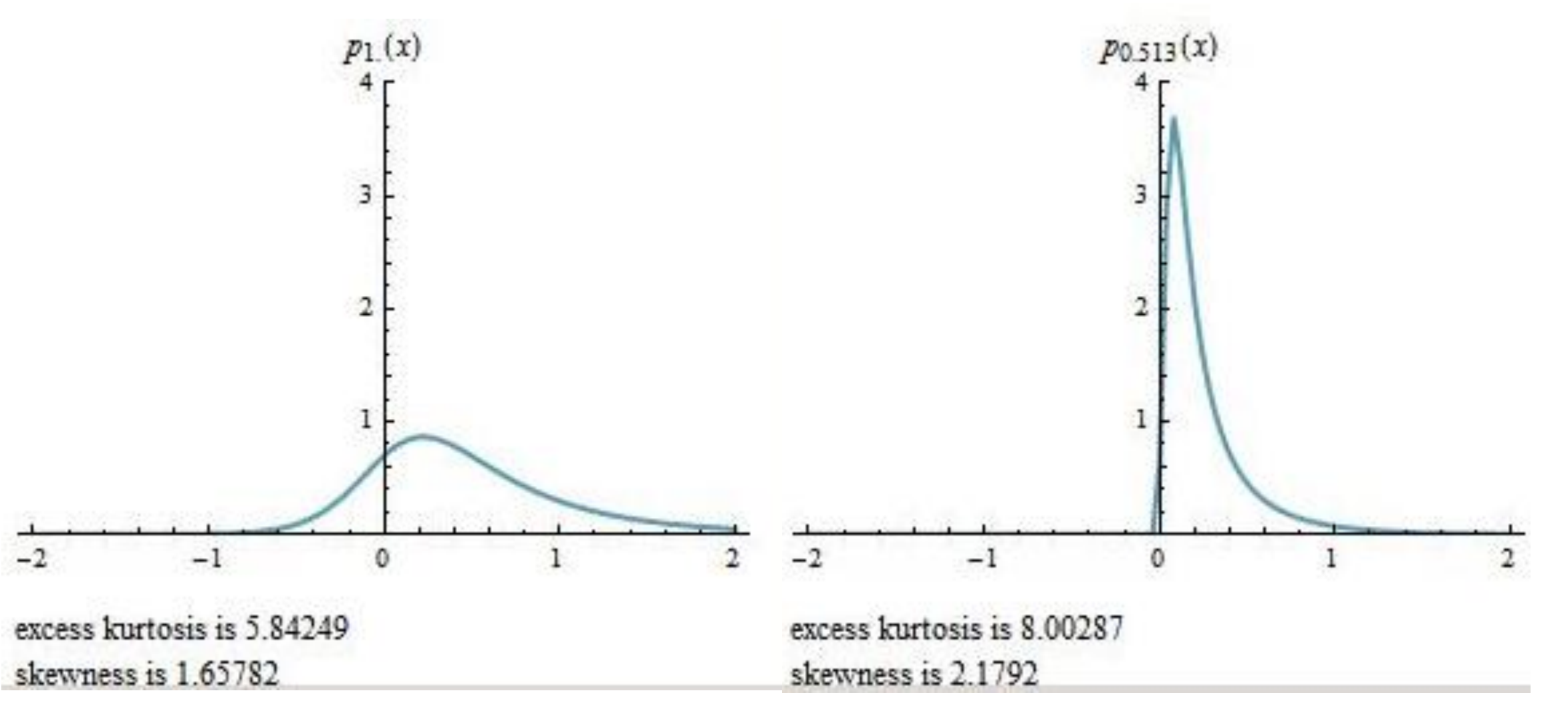

[17] Barndorff-Nielsen (1977), (1997), Stentoft (2008), Fragiadakis, Karlis, and Meintanis (2009), Eriksson, Ghysels and Wang (2009), Jönsson, Masol and Schoutens (2010)

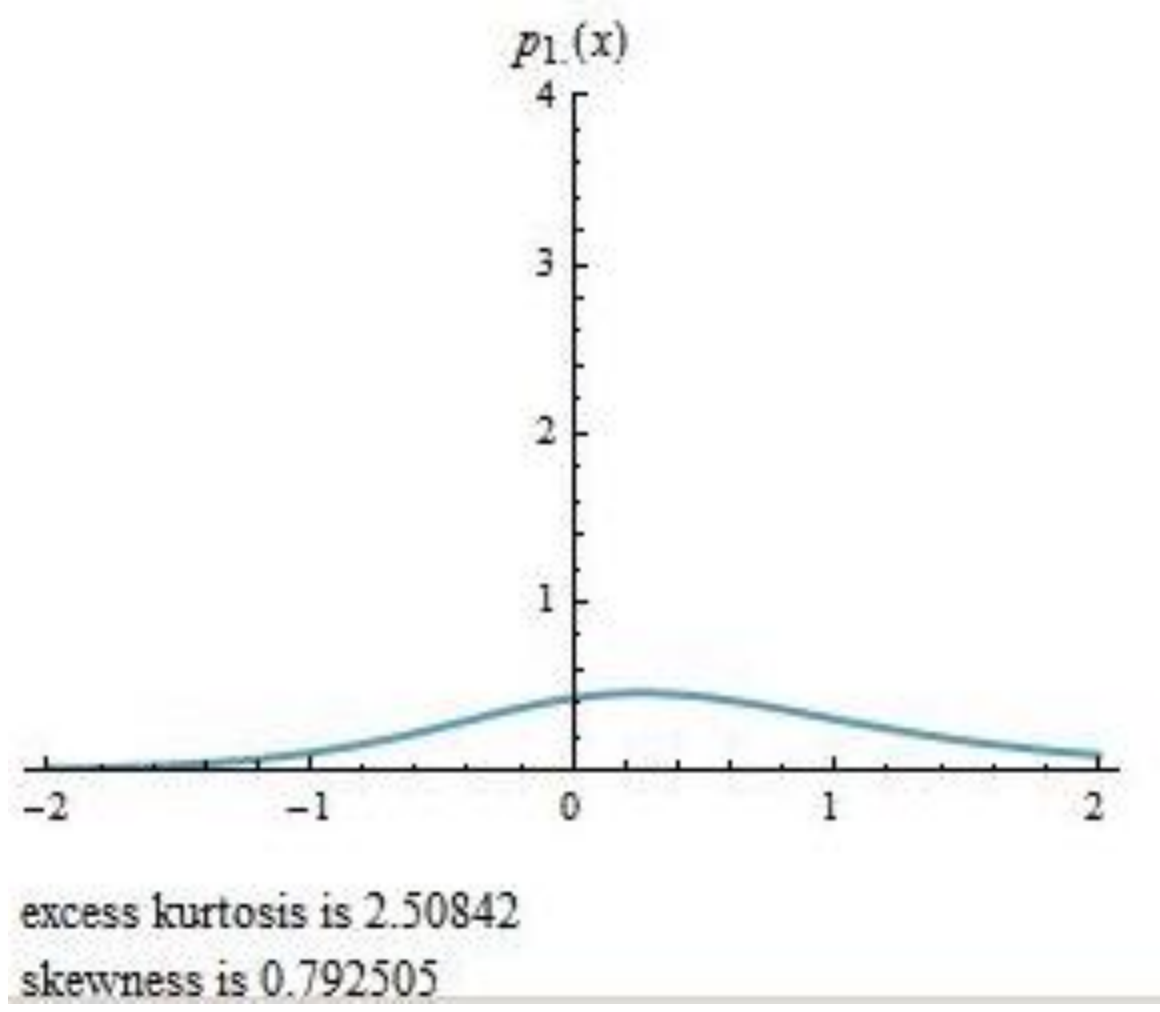

**Figure 1: Plots of NIG-probability density function (Wolfram Demonstrations Project)**

The NIG -distribution, $X \triangleq NIG(\mu, \alpha, \beta, \delta)$, encompasses the normal $\mathcal{N}(\mu, \sigma^2)$ distribution as a limiting case with $\beta = 0$, and $\sigma^2 = \lim_{\delta\uparrow\infty,\alpha\uparrow\infty} \frac{\delta}{\alpha}$. We shall now replace the log-normal assumption in Mehra and Prescott (1985) with log-NIG, and thus obtain Mehra and Prescott (1985) result as a limiting case. What is more important, is that by using log-NIG distribution the result will be flexible enough to give statistical- distributional explanation of the equity premium puzzle.

Let us briefly sketch Mehra-Prescott (1985) model. It assumes a frictionless economy with one representative investor ($\beth$ ) seeking to optimize the expected utility $\mathbb{E}_0(\sum_{t=0}^{\infty} \mathbb{b}^t \mathbb{U}(\mathbb{c}_t)$, where $\mathbb{b} \in (0,1)$is the discount factor, $\mathbb{U}(\mathbb{c}_t)$ is the utility from the consumption amount $\mathbb{c}_t$ at time $t = 0,1,2,\ldots$ The utility function is given by $\mathbb{U}(\mathbb{c}) = \mathbb{U}^{(\mathbb{a})}(\mathbb{c}) = \frac{\mathbb{c}^{1-\mathbb{a}}-1}{1-\mathbb{a}}$, where $\mathbb{a} > 0$ is the coefficient of relative risk aversion (CRRA) . $\beth$ invests in the asset at time $t$ giving $p_t$ units of consumption. $\beth$ sells the asset at $t+1$, receiving $p_{t+1} + y_{t+1}$, where $p_{t+1}$ is the asset price at $t+1$, and $p_{t+1}$ is the earned dividend at $t+1$. In the Mehra-Prescott model $\beth's$ return on investment in $(t, t+1]$ is given by

$$\mathbb{R}^{(e)}(t+1) = \frac{p_{t+1} + y_{t+1}}{p_t} = \mathbb{R}^{(f)}(t+1) - \frac{cov_t\left(\frac{\partial \mathbb{U}^{(\mathbb{a})}(\mathbb{c}_{t+1})}{\partial \mathbb{c}}, \mathbb{R}^{(e)}(t+1)\right)}{\mathbb{E}_t\left(\frac{\partial \mathbb{U}^{(\mathbb{a})}(\mathbb{c}_{t+1})}{\partial \mathbb{c}}\right)},$$

where $\mathbb{R}^{(f)}(t+1)$ is the riskless rate at $t+1$. Mehra and Prescott defined conception growth in $(t, t+1]$ as $\mathbb{x}_{t+1} = \frac{\mathbb{c}_{t+1}}{\mathbb{c}_t}$, which yields to

$$\mathbb{R}^{(e)}(t+1) = \frac{\mathbb{E}_t(\mathbb{x}_{t+1})}{\mathbb{b}\mathbb{E}_t(\mathbb{x}_{t+1}^{1-\mathbb{a}})}, \mathbb{R}^{(f)}(t+1) = \frac{1}{\mathbb{b}\mathbb{E}_t(\mathbb{x}_{t+1}^{-\mathbb{a}})}.$$

Mehra and Prescott assumes that the $\mathbb{x}_t, t = 1,2, \ldots$, are independent identically distributed with $\mathbb{x}_t \triangleq ln\mathcal{N}\left(\mu^{(x)}, {\sigma^{(x)}}^2\right)$ this leads to the following expression for the equity premium:

$$ln\mathbb{R}^{(e)}(t+1) - \ln \mathbb{R}^{(f)}(t+1) = \mathbb{a}{\sigma^{(x)}}^2. \qquad (8)$$

Thus, the equity premium is equal to $\beth's$ risk aversion times the variance of the consumption growth. Testing their model on US data for the period of 1889 to 1978 Mehra and Prescott find that $\mathbb{a}$ was close to 10, rather the general consensus estimate of $\mathbb{a}$ close to 3.[18]

We assume now that $ln\mathbb{x}_{t+1} \triangleq NIG(\mu^{(x)}, \alpha^{(x)}, \beta^{(x)}, \delta^{(x)})$[19]. From (7), it follows that

$$\mathbb{E}\left(\mathbb{R}^{(e)}(t+1)\right) = \frac{\exp\left\{\mu^{(x)} + \delta^{(x)}\left(\sqrt{{\alpha^{(x)}}^2 - {\beta^{(x)}}^2} - \sqrt{{\alpha^{(x)}}^2 - (\beta^{(x)}+1)^2}\right)\right\}}{\mathbb{b}\exp\left\{\mu^{(x)}(1-\mathbb{a}) + \delta^{(x)}\left(\sqrt{{\alpha^{(x)}}^2 - {\beta^{(x)}}^2} - \sqrt{{\alpha^{(x)}}^2 - (\beta^{(x)}+1-\mathbb{a})^2}\right)\right\}}$$

[18] See Mehra and Prescott (2003), Læssøe S., and Diernisse M. (2011).

[19] We shall use also with alternative notation $\mathbb{x}_{t+1} \triangleq lnNIG\left(\mu^{(x)}, \alpha^{(x)}, \beta^{(x)}, \delta^{(x)}\right)$.

and

$$\mathbb{R}^{(f)}(t+1) = \frac{1}{\mathbb{b}\exp\left\{\mu^{(x)}(-\mathbb{a})+\delta^{(x)}\left(\sqrt{{\alpha^{(x)}}^2-{\beta^{(x)}}^2}-\sqrt{{\alpha^{(x)}}^2-\left(\beta^{(x)}-\mathbb{a}\right)^2}\right)\right\}}.$$

Thus, we have the following extension of Mehra-Prescott equity premium:

$$ln\mathbb{E}\left(\mathbb{R}^{(e)}(t+1)\right) - ln\mathbb{R}^{(f)}(t+1) =$$

$$= \delta^{(x)}\left(\begin{array}{c} \sqrt{{\alpha^{(x)}}^2-{\beta^{(x)}}^2} - \sqrt{{\alpha^{(x)}}^2-(\beta^{(x)}-\mathbb{a})^2} - \\ -\sqrt{{\alpha^{(x)}}^2-(\beta^{(x)}+1)^2} + \sqrt{{\alpha^{(x)}}^2-(\beta^{(x)}+1-\mathbb{a})^2} \end{array}\right). \qquad (9)$$

When $\beta^{(x)} = 0, \delta^{(x)} = {\sigma^{(x)}}^2\alpha^{(x)}$, then $\alpha^{(x)} \uparrow \infty$,

$$ln\mathbb{E}\left(\mathbb{R}^{(e)}(t+1)\right) - ln\mathbb{R}^{(f)}(t+1)$$

$$= {\sigma^{(x)}}^2\alpha^{(x)}\left({\alpha^{(x)}}^2 - \sqrt{{\alpha^{(x)}}^2-\mathbb{a}^2} - \sqrt{{\alpha^{(x)}}^2-1} + \sqrt{{\alpha^{(x)}}^2-(1-\mathbb{a})^2}\right)$$

$$\to {\sigma^{(x)}}^2\mathbb{a}$$

That is, we obtain Mehra-Prescott's equity premium (8) as a limiting case of (9).

To compare (8), and (9), let us standardize (8) with $\sigma^{(x)} = 1$, and (9) with $\beta^{(x)} = 0$, and $\delta^{(1)} = 1$. Then $\mathbb{E}ln\mathbb{x}_{t+1} = \mu$, variance $varln\mathbb{x}_{t+1} = 1$, the skewness $\gamma(ln\mathbb{x}_{t+1}) = 0$, and excess kurtosis $\kappa(ln\mathbb{x}_{t+1}) = \frac{3}{|\alpha^{(x)}|}$. Consider than the ratio of the right-hand sides of (8) and (9):

$$R(\mathbb{a},\alpha^{(x)}) = \frac{\alpha^{(x)}\left(\alpha^{(x)} - \sqrt{{\alpha^{(x)}}^2-\mathbb{a}^2} - \sqrt{{\alpha^{(x)}}^2-1} + \sqrt{{\alpha^{(x)}}^2-(1-\mathbb{a})^2}\right)}{\mathbb{a}}$$

The following plot shows that in the neighborhood of $\mathbb{a}\sim 10$, and $\kappa(ln\mathbb{x}_{t+1}) = \frac{3}{|\alpha^{(x)}|}\sim 0.3$ the right side of (9) is about 3 times smaller than the right-hand side of (8), as desired.

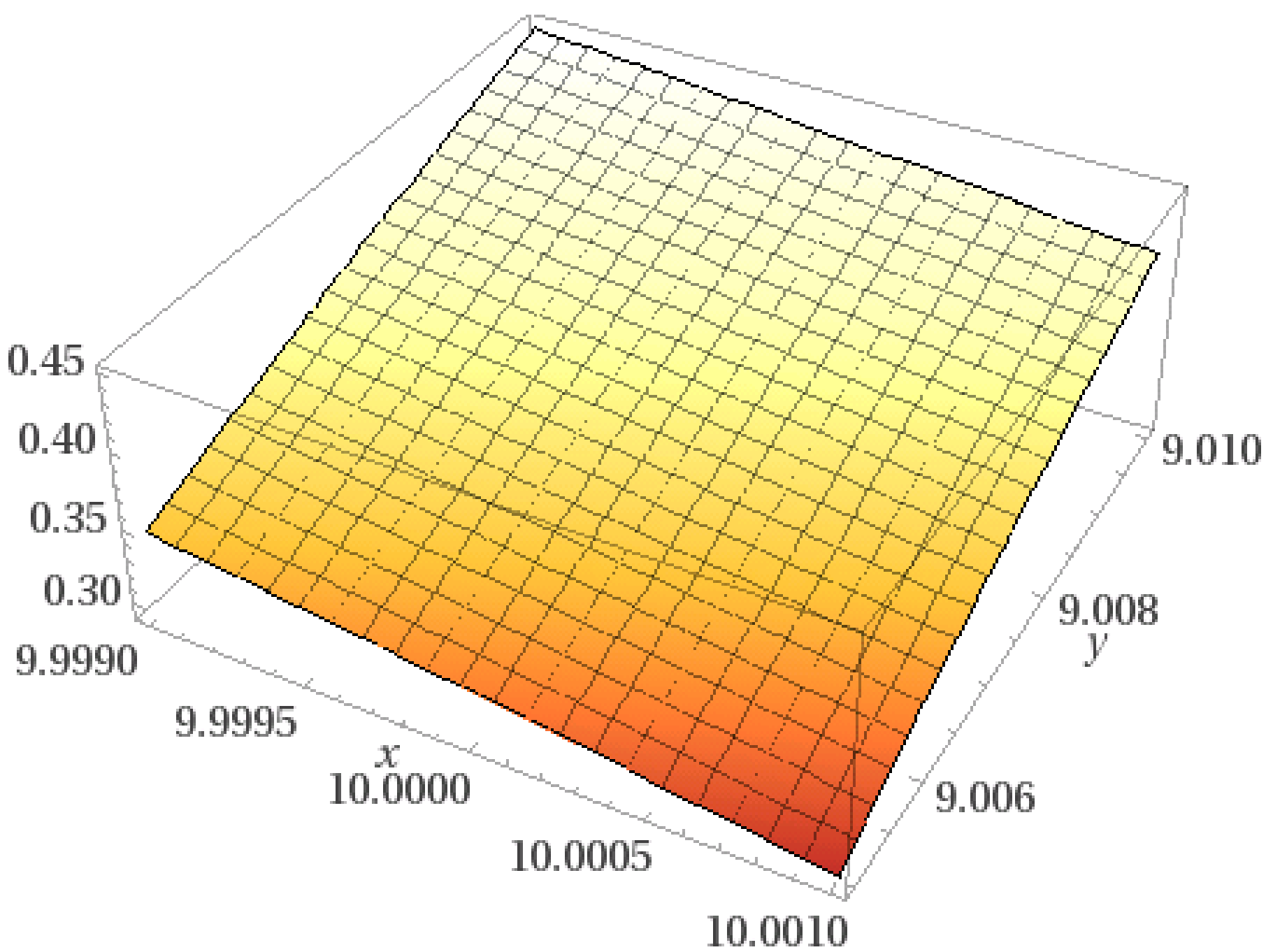

**Figure 2: Plot of $R(x,y), x = \mathbb{a} \in [9.999, 10.001], y = \alpha^{(x)} \in [9.005, 10.01]$.**

Our conclusion is that the equity -premium puzzle, as stated by Mehra and Prescott, can be explained, at least it terms of choosing more appropriate distribution for the consumption growth.

3. **Volatility Puzzle: The Rational Finance Approach**

Robert Shiller wrote[20]:” *The most significant market anomaly that efficient market theory fails to explain is excess volatility. The idea that stock prices change more than they rationally should is more troubling for efficient market theorists than any other anomaly, such as the January effect*

[20] Shiller (2003)

*or the day-of-the-week effect. If most of the volatility in the stock market is unexplained, then efficient market theory can be easily challenged. Efficient market theory says that asset prices can be forecast using the present discounted value of future returns. Yet because of excess volatility, forecasts of stock prices based on this idea tend to be more unreliable than the prices themselves. Some efficient market theorists argue that prices are efficient at the individual stock level but not at the aggregate market level, but others concede that the level of volatility in the overall stock market cannot be explained with any variant of the efficient market model."*

In term of volatility puzzle[21], a number of papers, most recently Santos and Woodford (1997), attempt to show the conditions under which rational bubbles[22] can survive are extremely restrictive[23]. Th behavioral finance proponents assert that investors believe that the mean dividend growth rate is more variable than it actually is. Similarly, price-dividend ratios and returns might also be excessively volatile because as behavioralists claim investors extrapolate past returns too far into the future when forming expectations of future returns[24].

---

[21] See Olsen (1998) , Thaler (1991) ,(1993) and Wood (1995).

[22] Brunnermeier (2001).

[23] Our opinion is that rational bubbles occur due to pre-critical conditions of the financial market as dynamical system, which over time lead to phase-transition, the bubble bursts, see for example, Fabretti and Ausloos (2004) , Yalamova and McKelvey (2011), Yukalov, Yukalova, and Sornette (2015)

[24] See the discussion sin Fisher (1928), Shafir, Diamond, and Tversky (1997 Ritter and Warr (2002), claim that the variation in P/D ratios and returns are due to investors mixing real and

We believe, resolutions of the volatility puzzle within the theory of rational finance already has been already proposed successfully[25]. We illustrate the rational finance approach to volatility puzzle, on the following extension of LeRoy-Lansing Model (LLM)[26].

In the LLM, $R_{t+1}^{(S)}, t = 0,1, \ldots$ is the stock gross return in the period $(t, t+1]$, and it has the representation

$$R_{t+1}^{(S)} = \frac{p_{t+1}^{(S)} + d_{t+1}}{p_t^{(S)}} = \left(\frac{z_{t+1}^{(S)}}{\mathbb{E}_t z_{t+1}^{(S)}}\right)\left(\frac{M_t}{M_{t+1}}\right),$$

where $p_t^{(S)}$ is the ex-dividend stock price at $t$, $\mathbb{E}_t z_{t+1}^{(S)} = \mathbb{E}\left(z_{t+1}^{(S)}/\mathcal{F}_t\right)$, $\mathcal{F}_t$ is the information until time $t$ , $d_{t+1}$ is the dividend received in $(t, t+1]$, $M_t$ is the stochastic discount factor at $(t, t+1]$ and $z_{t+1}^{(S)} \coloneqq \left(\frac{M_{t+1}}{M_t}\right)\left(\frac{d_{t+1}}{d_t}\right)\left(1 + \frac{p_{t+1}^{(S)}}{d_{t+1}}\right)$[27]. Similarly, in the LLM, the gross bond return $R_{t+1}^{(B)}$ in the period $(t, t+1]$, has the representation

$$R_{t+1}^{(B)} = \frac{1 + q p_{t+1}^{(B)}}{p_t^{(B)}} = \left(\frac{z_{t+1}^{(B)}}{\mathbb{E}_t z_{t+1}^{(B)}}\right)\left(\frac{M_t}{M_{t+1}}\right),$$

---

nominal quantities when forecasting future cash flows. Barberis, Huang and Santos (2001) show that the degree of loss aversion depends on prior gains and losses.

[25] See for example Kawakami (2016).

[26] LeRoy and Lansing (2016).

[27] LeRoy and Lansing (2016) (p.4) introduced $z_{t+1}^{(S)}$ as a composite variable that depends on the stochastic discount factor, the growth of dividends and the price-dividend ratio.

where $p_t^{(B)}$ is the price at $t$ of a default-free bond initiated at $R_{t+1}^{(S)} = 0$, $q^t$ $(0 < q < 1))$ is the coupon at time $t$, and $z_{t+1}^{(B)} \coloneqq \left(\frac{M_{t+1}}{M_t}\right)\left(1 + qp_{t+1}^{(B)}\right)$. Thus, the stock excess return has the form

$$lnR_{t+1}^{(S)} - lnR_{t+1}^{(B)} = ln\, z_{t+1}^{(S)} - ln\mathbb{E}_t z_{t+1}^{(S)} - ln\, z_{t+1}^{(B)} + ln\mathbb{E}_t z_{t+1}^{(B)}.$$

Next, in the LLM, it is assumed that conditional on $\mathcal{F}_t$, $z_{t+1/\mathcal{F}_t}^{(S)}$ and $z_{t+1/\mathcal{F}_t}^{(B)}$ are log-normally distributed, $z_{t+1/\mathcal{F}_t}^{(S)} \triangleq ln\mathcal{N}\left(\mu_t^{(S)}, {\sigma_t^{(S)}}^2\right)$ and , $z_{t+1//\mathcal{F}_t}^{(B)} \triangleq ln\mathcal{N}\left(\mu_t^{(B)}, {\sigma_t^{(B)}}^2\right)$. Thus,

$$lnR_{t+1}^{(S)} - lnR_{t+1}^{(B)} = ln\, z_{t+1//\mathcal{F}_t}^{(S)} - \left(\mu_t^{(S)} + \frac{1}{2}{\sigma_t^{(S)}}^2\right) - ln\, z_{t+1/\mathcal{F}_t}^{(B)} + \left(\mu_t^{(B)} + \frac{1}{2}{\sigma_t^{(B)}}^2\right),$$

and

$$\mathbb{E}_t lnR_{t+1}^{(S)} - \mathbb{E}_t lnR_{t+1}^{(B)} =$$

$$= \mathbb{E}_t\, ln\, z_{t+1//\mathcal{F}_t}^{(S)} - \left(\mu_t^{(S)} + \tfrac{1}{2}{\sigma_t^{(S)}}^2\right) - \mathbb{E}_t\, ln\, z_{t+1//\mathcal{F}_t}^{(B)} + \left(\mu_t^{(B)} + \tfrac{1}{2}{\sigma_t^{(B)}}^2\right) =$$

$$= -\tfrac{1}{2}{\sigma_t^{(S)}}^2 + \tfrac{1}{2}{\sigma_t^{(B)}}^2. \qquad (10)$$

Taking the unconditional variance in (10), leads to

$$var\left(\mathbb{E}_t lnR_{t+1}^{(S)} - \mathbb{E}_t lnR_{t+1}^{(B)}\right) = \tfrac{1}{2} var\left({\sigma_t^{(S)}}^2 - {\sigma_t^{(B)}}^2\right) =$$

$$= \tfrac{1}{2} var\left(var_t\left(ln\, z_{t+1//\mathcal{F}_t}^{(S)}\right) - var_t\left(ln\, z_{t+1//\mathcal{F}_t}^{(B)}\right)\right). \qquad (11)$$

Based on (11), LeRoy and Lansing (2016) (p.6) claim: *"The left-hand side of [(11)] gives a measure of the predictable variation in excess returns. Eq. [(11)] shows that if the conditional variances of [*$z_{t+1/\mathcal{F}_t}^{(S)}$*] and [* $z_{t+1/\mathcal{F}_t}^{(B)}$*] are constant across date-t events (although generally not equal to zero), then excess returns on stock are unpredictable. In that case markets are efficient.*

*If, on the other hand, the conditional variances differ according to the event, then a strictly positive fraction of excess returns are forecastable, so markets are inefficient."*[28]

We now extend LLM, assuming that $z_{t+1/\mathcal{F}_t}^{(S)}$ and $z_{t+1/\mathcal{F}_t}^{(B)}$ are log-NIG distributed $z_{t+1/\mathcal{F}_t}^{(S)} \triangleq lnNIG\left(\mu_t^{(S)}, \alpha_t^{(S)}, \beta_t^{(S)}, \delta_t^{(S)}\right)$, $z_{t+1/\mathcal{F}_t}^{(B)} \triangleq lnNIG\left(\mu_t^{(B)}, \alpha_t^{(B)}, \beta_t^{(B)}, \delta_t^{(B)}\right)$. Then,

$$lnR_{t+1}^{(S)} - lnR_{t+1}^{(B)} = ln\, z_{t+1}^{(S)} - \left(\mu_t^{(S)} + \delta_t^{(S)}\left(\sqrt{{\alpha_t^{(S)}}^2 - {\beta_t^{(S)}}^2} - \sqrt{{\alpha_t^{(S)}}^2 - \left(\beta_t^{(S)} - 1\right)^2}\right)\right) -$$

$$- ln\, z_{t+1}^{(B)} + \left(\mu_t^{(B)} + \delta_t^{(B)}\left(\sqrt{{\alpha_t^{(B)}}^2 - {\beta_t^{(B)}}^2} - \sqrt{{\alpha_t^{(B)}}^2 - \left(\beta_t^{(B)} - 1\right)^2}\right)\right).$$

---

[28] John Authers in his article "*Why are markets inefficient and what can be done about it?", in Financial Times of March 9, 2004, wrote "Markets are not perfectly efficient. More or less everyone agrees to this in the wake of the financial crisis. And while asset bubbles have recurred from time to time throughout history, bubble production has accelerated sharply. So not only are markets inefficient, but they are more inefficient than they used to be. This is despite rapid technological improvement to make markets faster and more liquid. So why are markets inefficient, and what can be done about it? The most popular answer is to blame human nature. Behavioral economists, applying experimental psychology, have explained many market anomalies. But human nature is constant. Greed and fear have been around forever. It is hard to blame an intensifying problem in the markets on any increased level of greed."* Our explanation is that the distribution of financial risk-factors are non-Gaussian distributed, as it will also be shown here, see next (13).

Then,

$$\mathbb{E}_t lnR_{t+1}^{(S)} - \mathbb{E}_t lnR_{t+1}^{(B)} =$$

$$= \left(\mu_t^{(S)} + \frac{\delta_t^{(S)}\beta_t^{(S)}}{\sqrt{{\alpha_t^{(S)}}^2-{\beta_t^{(S)}}^2}}\right) - \left(\mu_t^{(S)} + \delta_t^{(S)}\left(\sqrt{{\alpha_t^{(S)}}^2-{\beta_t^{(S)}}^2} - \sqrt{{\alpha_t^{(S)}}^2-\left(\beta_t^{(S)}-1\right)^2}\right)\right) -$$

$$- \left(\mu_t^{(B)} + \frac{\delta_t^{(B)}\beta_t^{(B)}}{\sqrt{{\alpha_t^{(B)}}^2-{\beta_t^{(B)}}^2}}\right) + \left(\mu_t^{(B)} + \delta_t^{(B)}\left(\sqrt{{\alpha_t^{(B)}}^2-{\beta_t^{(B)}}^2} - \sqrt{{\alpha_t^{(B)}}^2-\left(\beta_t^{(B)}-1\right)^2}\right)\right). \quad (12)$$

Taking the unconditional variance in (10), leads to

$$var\left(\mathbb{E}_t lnR_{t+1}^{(S)} - \mathbb{E}_t lnR_{t+1}^{(B)}\right) =$$

$$= var\left(\begin{array}{l} \frac{\delta_t^{(S)}\beta_t^{(S)}}{\sqrt{{\alpha_t^{(S)}}^2-{\beta_t^{(S)}}^2}} - \delta_t^{(S)}\left(\sqrt{{\alpha_t^{(S)}}^2-{\beta_t^{(S)}}^2} - \sqrt{{\alpha_t^{(S)}}^2-\left(\beta_t^{(S)}-1\right)^2}\right) - \\ -\frac{\delta_t^{(B)}\beta_t^{(B)}}{\sqrt{{\alpha_t^{(B)}}^2-{\beta_t^{(B)}}^2}} + \delta_t^{(B)}\left(\sqrt{{\alpha_t^{(B)}}^2-{\beta_t^{(B)}}^2} - \sqrt{{\alpha_t^{(B)}}^2-\left(\beta_t^{(B)}-1\right)^2}\right) \end{array}\right). \quad (13)$$

When $\beta_t^{(S)} = \beta_t^{(B)} = 0, \delta_t^{(S)} = {\sigma_t^{(S)}}^2\alpha_t^{(S)}, \delta_t^{(B)} = {\sigma_t^{(B)}}^2\alpha_t^{(B)}$ then $\alpha_t^{(S)} \uparrow \infty, \alpha_t^{(B)} \uparrow \infty,$ then (13) leads to (11). However, in general, we can adjust LeRoy and Lansing (2016) (p.6) claim as follows: *"The left-hand side of [(11)] gives a measure of the predictable variation in excess returns. Eq.(13) shows that if the tail indices* $\alpha_t^{(S)}, \alpha_t^{(B)}$, *the asymmetric parameters* $\beta_t^{(S)}, \beta_t^{(B)}$ *and the scale parameters* $\delta_t^{(S)}, \delta_t^{(B)}$ *of* $z_{t+1/\mathcal{F}_t}^{(S)}$ *and* $z_{t+1/\mathcal{F}_t}^{(B)}$ *] are constant across date-t events (although generally not equal to zero), then excess returns on stock are unpredictable. In that case markets are efficient. If, on the other hand, those parameters differ according to the event, more precisely, the right-hand side of (13) is time dependent, then a strictly positive fraction of excess returns are forecastable, so markets are inefficient."*

In conclusion, we claim that markets can often be inefficient when the factor returns have non-Gaussian distribution.

## 3. **Conclusion**

In this paper we concur with Mark Rubinstein[29] thesis: "*The Prime Directive:*

*Explain asset prices by rational models. Only if all attempts fail, resort to irrational investor behavior."* We study the three main objections of behavioral finance to the theory of rational finance, considered as "anomalies" the theory of rational finance cannot explain: (i) Predictability of asset returns; (ii) The Equity Premium; (iii) The Volatility Puzzle. We offer resolutions of those objections within the rational finance based on advanced techniques of modern finance, the use of Stratonovich integral in dynamic asset pricing (when addressing the (i) Predictability of asset returns), and special class of Lévy distributions (the normal inverse Gaussian distribution, the NIG-distribution) for asset returns when addressing the (ii) The Equity Premium; (iii) The Volatility Puzzle. Our conclusion is that modern asset pricing theory can explain the empirical; phenomena behavioral finance claim to exist in real financial market.

[29] Rubinstein (2000)